  \documentclass{cjour}
 \usepackage{cjourps,epsfig}
\newcommand{\vek}[1]{\mbox{\boldmath $   #1$}}

\begin{document}

%%%%% To be entered at Academic Press: =====>>
\finaltypesetting
% \journame{}
% \articlenumber{}
% \yearofpublication{}
% \volume{}
% \cccline{}
% \received{}
% \revised{}
% \accepted{}

\authorrunninghead{T. Winiecki and C. S. Adams}
\titlerunninghead{Semi-implicit finite difference method for the TDGL equations}

% communication line, use: \commline{Communicated by...}
% \commline{ }

%\setcounter{page}{261} %% This command is optional. 

%% <<== End of commands to be entered at Academic Press

%%  Authors, start here ==>>

%\draft % Optional, will cause a line at the bottom of each page
%% with the words `Draft' and the time and date that the article
%% was LaTeXed. Will also double space text.

\title{A Fast Semi-Implicit Finite Difference Method for the TDGL Equations}

%\subtitle{}

\author{T. Winiecki and C. S. Adams}
\affil{Dept. of Physics, University of Durham, Rochester Building,
South Road, Durham, DH1 3LE, England.}

%%%%%%%%%%%%
%% More than one author with separate affiliations, either:
%
%\author{First Author Name$^\dagger$ and Second Author Name$^\ddagger$}
%\affil{$^{\dagger}$First Author Affiliation, $^{\ddagger}Second Author
%       affiliation}

% or

% \author{Author name}
% \affil{Affiliation}
% \and
% \author{Author name}
% \affil{Affiliation}
%%%%%%%%%%%%

%% \thanks command:
%% Can use \thanks{} in title to have footnote number appear and
%% footnote at the bottom of the page. i.e.,
%% \title{This is the title\thanks{Supported by grant no....}}

%% In \authors or \affil, can use \thanks{} to have asterisk, 
%%   dagger or double dagger appear
%%   and text appear at the bottom of the title page. i.e.,

%\authors{D. Adalsteinsson and J. A. Sethian\thanks{Supported in part by the
%Applied Mathematics Subprogram of the...}}

%%%%%%%%%%%%

%\email{}

%optional
%\dedication{Dedicated to...}

\abstract{We propose a finite-difference algorithm for solving the time-dependent Ginzburg-Landau (TDGL) 
equation coupled to the appropriate Maxwell equation. The time derivatives are discretized using a
second order semi-implicit scheme which, for intermediate values of 
the Ginzburg-Landau parameter $\kappa$, allows time-steps two 
orders of magnitude larger than commonly used in explicit schemes.
We demonstrate the use of the method by solving a fully 
three-dimensional problem of a current-carrying wire with
longitudinal and transverse magnetic fields.
}

% text should be lower case, unless caps are necessary for meaning

\begin{article}

% \contents is optional, will make a list of all section heads 
% that appear in the article
%\contents

%optional
% for those that like to start with section zero:
%\zerosection{Introduction}

\section{Ginzburg-Landau Model}
In the Ginzburg-Landau model, a superconductor is characterised by a complex 
order parameter $\psi$. The local density of superconducting electrons is represented by 
$|\psi|^2$. The theory postulates that close to the critical temperature, the free
energy can be expanded in a series of the form
\begin{equation}
\begin{array}{ll}
{\cal L}(\psi,\nabla \psi,\vek{A},\nabla \times \vek{A})&=a|\psi|^2+\displaystyle{\frac{1}{2}}b|\psi|^4+
\displaystyle{\frac{\hbar^2}{2m_s}} \left |
\left( \nabla-{\rm i}\displaystyle{\frac{e_s}{\hbar}}\vek{A}\right) \psi \right |^2\\[12pt]
&+~\displaystyle{\frac{1}{2\mu_0}} \left | \nabla \times \vek{A}-\mu_0 \vek{H}\right|^2,
\end{array}
\end{equation}
where $a$ and $b$ are phenomenological parameters that depend on external parameters such as temperature,
$\vek{A}$ denotes the vector potential, $\vek{H}$ an external magnetic field, and $e_s$ and $m_s$ are the 
effective charge and the effective mass of the Cooper pairs.
Below the transition temperature $T_c$, $a$ becomes negative, whereas $b>0$ for all $T$. 
\subsection{The time-dependent Ginzburg-Landau (TDGL) equations}
The equations of motion for the order parameter and the vector potential are the Euler-Lagrange
equations of the free energy functional,
\begin{eqnarray}
\frac{\hbar^2}{2m_sD}\left( \partial_t + {\rm i} \frac{e_s}{\hbar} \Phi\right)\psi &=& 
\frac{\hbar^2}{2m_s}\left( \nabla - {\rm i} \frac{e_s}{\hbar}\vek{A} \right)^2\psi + |a|\psi-b|\psi|^2\psi
\label{eqn:TDGL_SI}\\
\frac{1}{\mu_0}\nabla \times (\nabla \times \vek{A}-\mu_0\vek{H})&=& \vek{j}_s+\vek{j}_n
\label{eqn:Maxwell_SI}\\
\vek{j}_s&=&\frac{\hbar e_s}{2m_s {\rm i}}(\psi^* \nabla \psi -\psi \nabla\psi^*)-\frac{e_s^2}{m_s}|\psi|^2 \vek{A}\\
\vek{j}_n&=&\sigma (-\nabla \Phi-\partial_t \vek{A})~,
\end{eqnarray}
where $D$ is a phenomenological diffusion constant, and
$\Phi$ is the electric potential, included to retain the gauge invariance of the equations.
Equation (\ref{eqn:Maxwell_SI}) is the Maxwell equation for the magnetic field, where the displacement current 
$\epsilon_0 \dot{\vek{E}}$ has been omitted as it only becomes significant 
for velocities close to the speed of light.
The total current is given by the sum of the supercurrent, $\vek{j}_s$, and the normal current, $\vek{j}_n$, 
which obeys Ohm's law.
\subsection{Dimensionless units}
We scale length in multiples of the coherence length, $\xi=\hbar/\sqrt{2m|a|}$, time in $\tau=\xi^2/D$, 
the wavefunction in $\psi_0=\sqrt{|a|/b}$, the vector potential in $A_0=\sqrt{2}\kappa H_c\xi$, 
where $H_c=\mu_0 |a|^2/b$, the electric potential
in $\Phi_0=(\xi/\tau)A_0$ and resistivity in units of the normal resistivity $\sigma_0=1/\kappa^2 D \mu_0$. 
The so-called Ginzburg-Landau parameter is given by
$\kappa^2=2 m^2 b/e^2\hbar^2\mu_0$. The characteristic length scale for variations of the magnetic field is 
$\lambda=\kappa \xi$ and $\nabla \times \vek{A}$ measures the magnetic field in 
units of $\sqrt{2}\kappa H_c=H_{c2}$. 
In scaled units equations (\ref{eqn:TDGL_SI}) and (\ref{eqn:Maxwell_SI}) become
\begin{eqnarray}
\left( \partial_t + {\rm i} \Phi\right)\psi &=& 
\left( \nabla - {\rm i} \vek{A} \right)^2\psi + \psi-|\psi|^2\psi
\label{eqn:TDGL_NoDim}\\
\kappa^2 \nabla \times \nabla \times \vek{A} &=& \underbrace{(\nabla S-\vek{A})|\psi|^2
}_{\vek{j}_s}+\underbrace{(-\nabla \Phi-\partial_t \vek{A})}_{\vek{j}_n} +\underbrace{\kappa^2 \nabla \times \vek{H}}_{\vek{j}_{ext}}~,
\label{eqn:AllCurrents}
\end{eqnarray}
where $S$ denotes the phase of $\psi$.
The last term in equation (\ref{eqn:AllCurrents}) can be understood as an external current $\vek{j}_{ext}$ with $\nabla \vek{j}_{ext}=0$.
In the following, this term will be omitted. However, it can be easily included in the algorithm,
for instance, to model magnetic impurities.
In dimensionless units, the dynamics of the superconductor depends on the dimensionless Ginzburg-Landau parameter
$\kappa$ only. For values $\kappa<1/\sqrt{2}$ one finds a behaviour characteristic of a type-I superconductor whereas 
for $\kappa>1/\sqrt{2}$ a type-II superconductor is modelled.
\subsection{Gauge transformation}
The dynamics of the measurable quantities $\vek{E}, \vek{B}, |\psi|^2$, and $\vek{j}$ are
invariant under the transformation
\begin{equation}
\left \{ \begin{array}{ll} 
\vek{A} &\to \vek{A}+ \nabla \Lambda\\ 
\psi &\to\psi {\rm e}^{{\rm i} \Lambda}\\
\Phi &\to\Phi - \dot{\Lambda} \end{array} \right.~,
\end{equation}
where $\Lambda$ is an arbitrary scalar field. We choose the zero potential gauge, 
$\Lambda(\vek{r},t)=\int {\rm d}t \Phi(\vek{r},t)$, in other
words, $\Phi(\vek{r}) \equiv 0$ at all times. For this choice, equations (\ref{eqn:TDGL_NoDim}) and (\ref{eqn:AllCurrents}) become
\begin{eqnarray}
\label{eqn:TDGL1}
\partial_t\psi &=& 
\left( \nabla - {\rm i} \vek{A} \right)^2\psi + \psi-|\psi|^2\psi\\
\label{eqn:TDGL2}
\partial_t \vek{A} &=& (\nabla S-\vek{A})|\psi|^2-\kappa^2 \nabla \times \nabla \times \vek{A}~.
\end{eqnarray}
In the following section, we suggest a fast and reliable numerical method to find an approximate solution 
to these equations.
\section{Numerical methods}
The most popular approach to the solution of the TDGL equations, (\ref{eqn:TDGL1}) and (\ref{eqn:TDGL2}), 
is a gauge-invariant discretization that is 
second order accurate in space and first order in time \cite{Grop1996,Frah1991,Kato1991,Masa1993,Crab2000}.
In addition, a number of other finite difference \cite{Cosk1997, Blac2000} and finite element methods
\cite{DuDu1994, DuDu1992} have been developed.
For large values of $\kappa$, the magnetic field is nearly homogeneous and equation (\ref{eqn:TDGL2}) can be dropped. 
This case is often referred to as the London limit. 
The remaining equation has been solved 
by a semi-implicit Fourier spectral method which is second order accurate in time \cite{Chen1998}. An equation very similar
to equation (\ref{eqn:TDGL1}), the Gross-Pitaevskii equation, is used to model vortex dynamics in dilute Bose-Einstein 
condensates \cite{Wini2000}. 
Here, we modify the very robust and accurate semi-implicit Crank-Nicholson algorithm used in \cite{Wini2000} to include 
the equation for the vector potential.
\subsection{The $U-\psi$ method}
The widely used $U-\psi$ method is described in detail by Gropp et. al. \cite{Grop1996}.
As this method forms the basis of our algorithm we briefly review the main points here.
Complex link variables $U^x$, $U^y$ and $U^z$ are introduced to preserve the gauge 
invariant properties of the discretized equations. 
\begin{equation}
\begin{array}{ll}
%{}&U^x(x,y,z)=\exp \left(-{\rm i} \int_{x_0}^x A^x(x',y,z) {\rm d} x' \right)\\
%{}&U^y(x,y,z)=\exp \left(-{\rm i} \int_{y_0}^y A^y(x,y',z) {\rm d} y' \right)\\
%{}&U^z(x,y,z)=\exp \left(-{\rm i} \int_{z_0}^z A^z(x,y,z') {\rm d} z' \right)~,
{}&\displaystyle{U^x(x,y,z)=\exp \left(-{\rm i} \int_{x_0}^x A^x(x',y,z) {\rm d} x' \right)}\\[12pt]
{}&\displaystyle{U^y(x,y,z)=\exp \left(-{\rm i} \int_{y_0}^y A^y(x,y',z) {\rm d} y' \right)}\\[12pt]
{}&\displaystyle{U^z(x,y,z)=\exp \left(-{\rm i} \int_{z_0}^z A^z(x,y,z') {\rm d} z' \right)~,}
\end{array}
\end{equation}
where $(x_0,y_0,z_0)$ is an arbitrary reference point.
The TDGL equations can then be expressed as functions of $\psi$ and these link variables. Both the order parameter 
and the link variables are discretized on a three dimensional grid with grid spacing $h_x$, $h_y$, and $h_z$, respectively.
The mesh points for the link variables
are half way between the mesh points for the order parameter (see Fig. \ref{fig:grid}).
All spatial derivatives are 
approximated by finite differences to second order accuracy. Denoting the complex conjugate of $U$ by $\overline{U}$, the 
finite difference representations of the TDGL equations read
\begin{eqnarray}
\nonumber
\partial_t\psi_{i,j,k}&=&
   \frac{\overline{U}^x_{i-1,j,k}\psi_{i-1,j,k}-2\psi_{i,j,k}+U^x_{i,j,k}\psi_{i+1,j,k}}{h_x^2}\\
\nonumber
&+&\frac{\overline{U}^y_{i,j-1,k}\psi_{i,j-1,k}-2\psi_{i,j,k}+U^y_{i,j,k}\psi_{i,j+1,k}}{h_y^2}\\
\nonumber
&+&\frac{\overline{U}^z_{i,j,k-1}\psi_{i,j,k-1}-2\psi_{i,j,k}+U^z_{i,j,k}\psi_{i,j,k+1}}{h_z^2}\\
&+&(1-|\psi_{i,j,k}|^2)\psi_{i,j,k}
\label{eqn:explicit1}\\
\partial_t U^x_{i,j,k}&=&-{\rm i}~{\rm Im}\left({\cal F}^x_{i,j,k}\right) U^x_{i,j,k}~,
\label{eqn:explicit2}
\end{eqnarray}
where
$$
\begin{array}{ll}
{\cal F}^x_{i,j,k}&=~\kappa^2 \displaystyle{\frac{\overline{U}^x_{i,j+1,k}\overline{U}^y_{i,j,k} U^x_{i,j,k} U^y_{i+1,j,k}  -  
            \overline{U}^x_{i,j,k}\overline{U}^y_{i,j-1,k} U^x_{i,j-1,k} U^y_{i+1,j-1,k}}{h_y^2}}\\[8pt]
&+~~\kappa^2 \displaystyle{\frac{\overline{U}^z_{i+1,j,k-1}\overline{U}^x_{i,j,k-1} U^z_{i,j,k-1} U^x_{i,j,k}-
            \overline{U}^z_{i+1,j,k}\overline{U}^x_{i,j,k} U^z_{i,j,k} U^x_{i,j,k+1}}{h_z^2}}\\[8pt]
&+~~U^x_{i,j,k}\overline{\psi}_{i,j,k}\psi_{i+1,j,k}~.
\end{array}
$$
Analogous expressions for $\partial_t U^y_{i,j,k}$ and $\partial_t U^z_{i,j,k}$ can be obtained by permutating the 
coordinates and indices as follows,
\begin{equation}
(x,y,z;i,j,k) \to (y,z,x;j,k,i) \to (z,x,y;k,i,j) \to (x,y,z;i,j,k)~.
\label{eqn:cyclic}
\end{equation}
The time evolution is approximated by a simple Euler step,
\begin{eqnarray}
\psi_{i,j,k}(t+\Delta t)&=&\psi_{i,j,k}(t)+\Delta t~\partial_t \psi_{i,j,k}(t)+ {\cal O}(\Delta t^2)\\
\label{eqn:Euler2}
U^x_{i,j,k}(t+\Delta t)&=&U^x_{i,j,k}(t)+\Delta t~\partial_t U^x_{i,j,k}(t)+ {\cal O}(\Delta t^2)~.
\end{eqnarray}
To keep $U^x_{i,j,k}$ uni-modular, equation (\ref{eqn:Euler2}) is often modified to
\begin{equation}
U^x_{i,j,k}(t+\Delta t)=U^x_{i,j,k}(t) \exp \left( - {\rm i} \Delta t~{\rm Im} {\cal F}^x_{i,j,k}\right)+{\cal O}(\Delta t^2)~.
\end{equation}
The Euler method is only first order accurate in time, i.e., the truncation error made due to the finite difference 
approximation of the time derivative is proportional to $\Delta t^2$. However, the main problem is that the code 
becomes unstable if long time steps are used.
The cause of this instability is the diffusion-like character of the dynamics described by the equations 
(\ref{eqn:explicit1}) and (\ref{eqn:explicit2}). Equation (\ref{eqn:explicit1}) 
can immediately be written as a diffusion equation with an additional non-linear term
\begin{equation}
\partial_t\psi_{i,j,k}= L_x \psi_{i,j,k}+L_y \psi_{i,j,k} +L_z \psi_{i,j,k} +f~,
\end{equation}
where $f$ stands for $(1-|\psi_{i,j,k}|^2)\psi_{i,j,k}$ and $L_x$, $L_y$, and $L_z$ denote the {\em weighted}
Laplacian operators,
\begin{equation}
L_x \psi_{i,j,k} \equiv \frac{a_{i-1}\psi_{i-1,j,k}-2\psi_{i,j,k}+a_{i+1}\psi_{i+1,j,k}}{h_x^2}~,
\end{equation}
with $|a_{i-1}|=|a_{i+1}|=1$ in our case. The diffusion constant is $1$ in dimensionless units. Equation
(\ref{eqn:explicit2}) is also dominated by diffusive terms as will become evident in the next section.
The diffusion constant for the vector potential is $\kappa^2$.
This can be seen by taking the curl of equation (\ref{eqn:TDGL2}),
$
\dot{\vek{B}}=\kappa^2\nabla^2 \vek{B}+\nabla \times \vek{j}_s.
$
The one-step forward Euler method is only stable as long as the time step is shorter than the diffusion time across a cell
of width $h$ \cite{NumRec}. For example, using a grid spacing of $h=0.5\xi$ and $\kappa=4$, the theoretical limit for the
time step is
\begin{equation}
\Delta t < \frac{h^2}{2\kappa^2}=\frac{0.5^2}{2 \cdot 4^2}\approx 0.0078~.
\end{equation}
In practice, a time step of $\Delta t=0.0025$ is used to ensure stability \cite{Grop1996}.
In contrast, a semi-implicit two-step algorithm is unconditionally stable for diffusive problems and enables
much larger time-steps to be employed.
\subsection{Semi-implicit algorithm}
\begin{figure}[ht]
\center
\epsfig{file=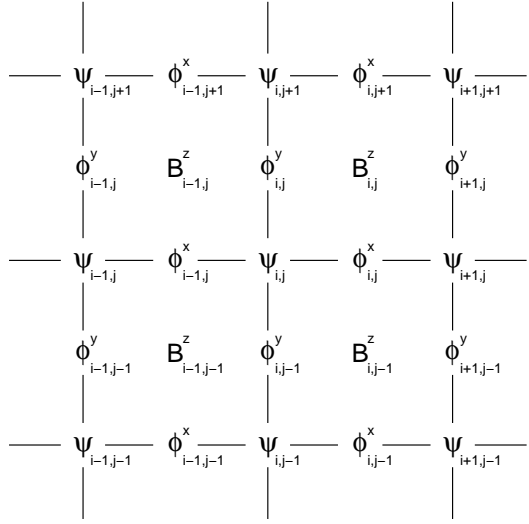,clip=,width=7cm}
\vskip12pt
\caption{The evaluation points for the fields $\psi$ and $\phi$ in the $x-y$ plane. A finite difference approximation for the magnetic
field $B_z$ is given in (\ref{eqn:BField}).}
\label{fig:grid}
\end{figure}

We propose a spatial discretization of the equations very similar to the above $U-\psi$ method.
The link variables are uni-modular, $|U^x_{i,j,k}|=1$, and can be written
as the exponential of a phase, $U^x_{i,j,k}=\exp(-{\rm i}\phi^x_{i,j,k})$.
We use the real-valued variable $\phi^x$ instead of the complex-valued $U^x$.
The fields $\psi$ and $\vek{\phi}$ are represented on a three-dimensional grid.
The mesh points of the phase factors 
are placed between the mesh points of the order parameter (see Fig. \ref{fig:grid}). For the field $\psi_{i,j,k}$,
the grid point indices are $i=1...N_x+1$, $j=1...N_y+1$, and $k=1...N_z+1$. For  $\phi^x_{i,j,k}$,
the indices in the $x$ direction range $i=1...N_x$ only, due to the relative displacement of the grids.
Similarly, $j=1...N_y$ for $\phi^y_{i,j,k}$ and $k=1...N_z$ for $\phi^z_{i,j,k}$.
\par We now discretize the spatial derivatives in equations (\ref{eqn:TDGL1}) and (\ref{eqn:TDGL2}) using the 
modified link variables $\phi^x$, $\phi^y$, and $\phi^z$. For equation (\ref{eqn:TDGL1}), we can reuse the 
expansion (\ref{eqn:explicit1}) except that $U^x_{i,j,k}$ is replaced by $\exp(-{\rm i}\phi^x_{i,j,k})$, etc,
\begin{equation}
\begin{array}{ll}
\partial_t \psi_{i,j,k}&=\displaystyle{\frac{\exp({\rm i}\phi^x_{i-1,j,k})\psi_{i-1,j,k}-2\psi_{i,j,k}+
\exp(-{\rm i}\phi^x_{i,j,k})\psi_{i+1,j,k}}{h_x^2}}\\[12pt]
&+~\displaystyle{\frac{\exp({\rm i}\phi^y_{i,j-1,k})\psi_{i,j-1,k}-2\psi_{i,j,k}+
\exp(-{\rm i}\phi^y_{i,j,k})\psi_{i,j+1,k}}{h_y^2}}\\[12pt]
&+~\displaystyle{\frac{\exp({\rm i}\phi^z_{i,j,k-1})\psi_{i,j,k-1}-2\psi_{i,j,k}+
\exp(-{\rm i}\phi^z_{i,j,k})\psi_{i,j,k+1}}{h_z^2}}\\[12pt]
&+~(1-|\psi_{i,j,k}|^2)\psi_{i,j,k}~.
\end{array}
\label{eqn:newFinDiff1}
\end{equation}
With help of the relation $-\nabla \times \nabla \times \vek{A}=\nabla^2 \vek{A} -\nabla(\nabla \vek{A})$, the
second order accurate finite difference representation of (\ref{eqn:TDGL2}) is

\begin{eqnarray}
\nonumber
\partial_t \phi^x_{i,j,k}&=&\frac{\kappa^2}{h_y^2}(\phi^x_{i,j+1,k}-2\phi^x_{i,j,k}+\phi^x_{i,j-1,k})+
                           \frac{\kappa^2}{h_z^2}(\phi^x_{i,j,k+1}-2\phi^x_{i,j,k}+\phi^x_{i,j,k-1})\\
\nonumber
&+&\frac{\kappa^2}{h_y^2}(-\phi^y_{i+1,j,k}+\phi^y_{i,j,k}+\phi^y_{i+1,j-1,k}-\phi^y_{i,j-1,k})\\
\nonumber
&+&\frac{\kappa^2}{h_z^2}(-\phi^z_{i+1,j,k}+\phi^z_{i,j,k}+\phi^z_{i+1,j,k-1}-\phi^z_{i,j,k-1})\\
&+&{\rm Im} \left( \exp(-{\rm i}\phi^x_{i,j,k})\overline{\psi}_{i,j,k}\psi_{i+1,j,k} \right)~.
\label{eqn:newFinDiff2}
\end{eqnarray}
The expressions for $\partial_t \phi^y_{i,j,k}$ and $\partial_t \phi^z_{i,j,k}$ are given by cyclic permutation 
(\ref{eqn:cyclic}).

Note, that the discretized equations are still invariant under the gauge transformation
\begin{equation}
\left \{ \begin{array}{ll} 
\psi_{i,j,k}&\to\psi_{i,j,k} \exp({\rm i}\Lambda_{i,j,k})\\
\phi^x_{i,j,k}&\to\phi^x_{i,j,k} +(\Lambda_{i+1,j,k}-\Lambda_{i,j,k})\\
\phi^y_{i,j,k}&\to\phi^y_{i,j,k} +(\Lambda_{i,j+1,k}-\Lambda_{i,j,k})\\
\phi^z_{i,j,k}&\to\phi^z_{i,j,k} +(\Lambda_{i,j,k+1}-\Lambda_{i,j,k})
\end{array} \right.~.
\end{equation}
Retaining the gauge invariance at the discrete level is often equivalent to preserving certain conservation
laws and physical principles. It is crucial that the numerical approximation does not
depend on the particular choice of gauge. If, for example, one studies the motion of a vortex
lattice due to an applied electric field $E_x$, the measurable quantities $\vek{B}$, $|\psi|^2$ and $\vek{j}$
oscillate in time \cite{Wini2001}. The system is driven through a series of equivalent solutions and the dynamics
is roughly described by $\Lambda = E_x xt$. This means that the phase gradients in the order parameter build up
in time and the phase difference between two neighbouring grid points eventually exceeds $2\pi$. This is normally a
problem as the finite difference approximation becomes invalid. However, using the 
link variables $U$ or $\phi$ these phase gradients are exactly cancelled by the change in the vector potential.

We now want to introduce a new scheme to update the wavefunction $\psi^{(n)}$and the link variables 
$\phi^{(n)}$ from the $n$th to the $(n+1)$th step.
The idea is to treat the diffusive terms semi-implicitly whereas all other terms are still treated explicitly.
In this way we reduce the stability constraints associated with the simple Euler method but avoid the expensive
solution of non-linear equations. The technique is known as {\it the method of fractional steps} \cite{Ames1992}.
A second-order accuracy in the time-step can 
be achieved by a simple 3 step iteration.

As mentioned above, the diffusive character of equation (\ref{eqn:TDGL2}) becomes apparent in the new discretization
and both equation (\ref{eqn:newFinDiff1}) and equation (\ref{eqn:newFinDiff2}) can be written as a initial value 
problem of the form
\begin{equation}
\partial_tu_{i,j,k}= D \left(L_x u_{i,j,k}+L_y u_{i,j,k} +L_z u_{i,j,k} \right) +f~,
\label{eqn:ut}
\end{equation}
where $u$ stands for the fields $\psi$ or $\phi^x$, $\phi^y$ or $\phi^z$, respectively, $D$ is the diffusion 
constant with $D=\kappa^2$ in  (\ref{eqn:newFinDiff2}) and $f$ indicates all the other terms, $(1-|\psi_{i,j,k}|^2)\psi_{i,j,k}$ in 
(\ref{eqn:newFinDiff1}) and the last three lines in equation (\ref{eqn:newFinDiff2}).
Note, that $L_x \equiv 0$ in (\ref{eqn:newFinDiff2}).

The second derivatives are approximated in the usual way by an expression involving three neighbouring grid points. For
any pair $(j,k)$ the action of $L_x$ on the vector $\{u_{i,j,k}\}_{i=2...N_x}$,  can be represented
by a tri-diagonal matrix $\delta_x^2$,
\begin{equation}
L_x u_{i,j,k} \equiv \frac{\delta_x^2}{h_x^2}u_{i,j,k}\equiv \frac{a_{i-1} u_{i-1,j,k}-2u_{i,j,k}+a_{i+1}u_{i+1,j,k}}{h_x^2} ~.
\end{equation}
As emphasized before, the instabilities of the Euler method
have their origin in the explicit treatment of the diffusive terms.
We now discretize the time derivative in equation (\ref{eqn:ut}) in the following way,
\begin{equation}
\begin{array}{ll}
\displaystyle{\frac{u^{(n+1)}-u^{(n)}}{\Delta t}}&=\displaystyle{\frac{1}{2}}\displaystyle{\frac{D\delta_{x}^2}{h_x^2}}\left(u^{(n+1)}+u^{(n)}\right)+
\displaystyle{\frac{1}{2}}\displaystyle{\frac{D\delta_{y}^2}{h_y^2}}\left(u^{(n+1)}+u^{(n)}\right)\\[12pt] {}&+
\displaystyle{\frac{1}{2}}\displaystyle{\frac{D\delta_{z}^2}{h_z^2}}\left(u^{(n+1)}+u^{(n)}\right)+\displaystyle{\frac{1}{2}}\left( f^{(n+1)}+f^{(n)}\right)+{\cal O}(\Delta t^2)~.
\end{array}
\end{equation}
This discretization is semi-implicit as the right 
hand side of the equation depends on the fields at the old and the new time level. This mixing leads to an 
improved accuracy and prevents the algorithm from developing instabilities.
After rearranging the equation we get
\begin{equation}
\begin{array}{ll}
{}&\left( 1-\displaystyle{\frac{D\Delta t}{2h_x^2}\delta_{x}^2}-\displaystyle{\frac{D\Delta t}{2h_y^2}\delta_{y}^2}-\displaystyle{\frac{D\Delta t}{2h_z^2}\delta_{z}^2} \right) u^{(n+1)}=\\[12pt]
{}&\left( 1+\displaystyle{\frac{D\Delta t}{2h_x^2}\delta_{x}^2}+\displaystyle{\frac{D\Delta t}{2h_y^2}\delta_{y}^2}+\displaystyle{\frac{D\Delta t}{2h_z^2}\delta_{z}^2} \right) u^{(n)}
+\displaystyle{\frac{\Delta t}{2}}\left( f^{(n+1)}+f^{(n)}\right)+{\cal O}(\Delta t^3)~.
\end{array}
\label{eqn:notfactorized}
\end{equation}
We now employ an {\em approximate factorisation} \cite{Ames1992},
\begin{equation}
\begin{array}{ll}
{}&\left( 1-\displaystyle{\frac{D\Delta t}{2h_x^2}}\delta_{x}^2-\displaystyle{\frac{D\Delta t}{2h_y^2}}\delta_{y}^2-\displaystyle{\frac{D\Delta t}{2h_z^2}}\delta_{z}^2 \right) =\\[12pt]
{}&\left( 1-\displaystyle{\frac{D\Delta t}{2h_x^2}}\delta_{x}^2\right)
\left( 1-\displaystyle{\frac{D\Delta t}{2h_y^2}}\delta_{y}^2\right)
\left( 1-\displaystyle{\frac{D\Delta t}{2h_z^2}}\delta_{z}^2\right)+{\cal O}(\Delta t^2)~.
\end{array}
\end{equation}
The multidimensional operator is split into three operators that involve difference approximations in only one 
dimension. With the abbreviations
\begin{equation}
A_x=\left(1 - \frac{D\Delta t}{2h_x^2}\delta_{x}^2\right), 
~B_x=\left(1 + \frac{D\Delta t}{2h_x^2}\delta_{x}^2\right),
\end{equation}
and considering $(u^{(n+1)}-u^{(n)})={\cal O}(\Delta t)$ equation (\ref{eqn:notfactorized}) becomes
\begin{equation}
A_x A_y A_z u^{(n+1)}=
B_x B_y B_z u^{(n)}+\frac{\Delta t}{2}\left( f^{(n+1)}+f^{(n)}\right)+{\cal O}(\Delta t^3)~.
\end{equation}
The tri-diagonal matrices $A$ and $B$ are actually time dependent because the differential operators $L$ in 
equation (\ref{eqn:newFinDiff1}) depend on the link variables. In the above equation,
$A$ is a function of $\phi^{(n+1)}$ whereas $B$ depends on $\phi^{(n)}$. 
Consequently, the equations are solved in the following step-wise manner
\begin{equation}
\begin{array}{ll}
\displaystyle{A_x u^{(n+1/3)}}&\displaystyle{=~B_x B_y B_z u^{(n)} + \frac{\Delta t}{2} \left(f^{(n+1)}+f^{(n)}\right)}\\[6pt]
\displaystyle{A_y u^{(n+2/3)}}&\displaystyle{=~u^{(n+1/3)}}\\[6pt]
\displaystyle{A_z u^{(n+1)}}&\displaystyle{=~u^{(n+2/3)}}~,
\end{array}
\label{eqn:solveTridiag}
\end{equation}
where the `fractional' time levels indicate intermediate results.
The explicit term $f^{(n+1)}$ as well as the matrix elements of $A$ may depend on the values of the link variables 
at the new time level and are unknown initially. We 
assume that $u^{(n+1)}=u^{(n)}$ to start with. After the first iteration of equation (\ref{eqn:solveTridiag}) for
{\em all} variables $\psi$ and $\phi$, the updated values at the new time level are used in the matrix elements of $A$
for the second iteration, and so on.
The product $B_x B_y B_z u^{(n)}$ is a function of known values at the previous time level and can be stored in 
an auxiliary variable for subsequent iterations. As the matrices $A$ are tri-diagonal, fast inversion routines
can be applied \cite{NumRec}.
\par
\begin{figure}[ht]
\center
\epsfig{file=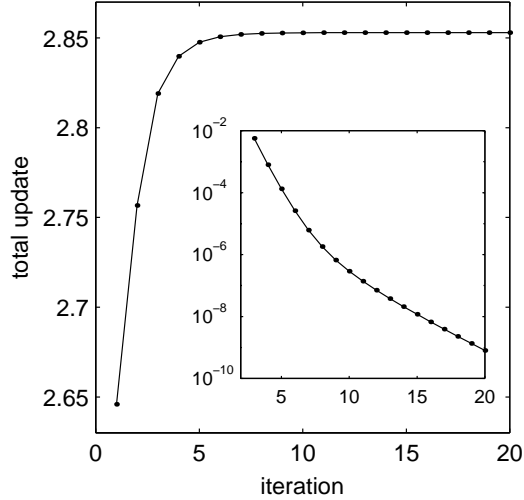,clip=,width=7cm}
\vskip12pt
\caption{The modification to the solution after one time step (total update) versus the number 
of iterations. For smaller time-steps, the solution converges faster. 
We find an optimum of speed and accuracy for a combination of three iterations and a time-step of $\Delta t=0.5$.
The correction between iterations, $T(m)$ is shown inset.}
\label{fig:iterations}
\end{figure}

 The entire algorithm relies on the convergence of this iteration technique. To test whether the procedure 
converges, we calculate a total update, $S(m)$, of the fields after $m$ iterations by comparing all values 
at the time level $(n+1)$ to all values at the previous time step, $(n)$, 
\begin{equation}
S(m)=\sum_{i,j,k} \left( \left ( \left|\psi_{i,j,k}^{(n+1,m)}\right|^2-\left|\psi_{i,j,k}^{(n)}\right|^2\right)^2 +
\left ( \phi_{i,j,k}^{x(n+1,m)}-\phi_{i,j,k}^{x(n)}\right)^2+...\right)~,
\end{equation}
where the three dots indicate the corresponding terms for the fields $\phi^y$ and
$\phi^z$.
Fig. \ref{fig:iterations} shows a typical evolution of the update for a time step of $\Delta t=0.5$.
After as few as five iterations the approximated increment is very close to the exact value. For smaller time steps, the procedure
converges faster. We find an optimum trade-off between accuracy and performance for three iterations. 
We further check, if the correction between two successive iterations, 
\begin{equation}
\begin{array}{ll}
\displaystyle{T(m)=\sum_{i,j,k}}&\left( \left ( \left|\psi_{i,j,k}^{(n+1,m)}\right|^2-\left|\psi_{i,j,k}^{(n+1,m-1)}\right|^2\right)^2 \right.\\[10pt]
{}&+ \left.\left ( \phi_{i,j,k}^{x(n+1,m)}-\phi_{i,j,k}^{x(n+1,m-1)}\right)^2+...\right)
\end{array}
\end{equation}
converges to zero. Fig. (\ref{fig:iterations}, inset) confirms an exponential convergence.
\par The accuracy of the method is assessed by comparing the solution to
simulations using the Euler method with a much smaller time step, $\Delta t=0.0025$.
Up to a time step of $\Delta t=0.5$, no significant deviations could be observed. The program runs about 40 times
faster than the Euler method for these parameters. For the calculations below we use a finer grid ($h=0.4$) and a slightly larger
Ginzburg-Landau parameter ($\kappa=5$). The speed-up for these values is about 100.
\par Our implicit method is less memory intensive than the standard $U-\psi$ method because it uses real-valued
link variables rather than complex-valued ones that must be represented by two real numbers.
For a grid of $N^3$ points, the Euler method uses an equivalent of $22N^3$ real-valued variables (the complex
wavefunction
and the three complex link variables at two time levels plus three complex fields $W$, see \cite{Grop1996}) whereas the 
implicit method requires the storage of a total of $19N^3$ variables (the complex wavefunction and the three real link variables 
at two time levels, the products $B_xB_yB_z u^{(n)}$ in (\ref{eqn:solveTridiag}) plus four auxiliary fields).
\subsection{Boundary conditions}
The correct implementation of the boundary conditions requires great care because of the relative displacement of 
the grids.
The matrices $A_x$ and $B_x$ only act on the interior points, $i=2...N_x$, of the vectors
$\psi_{i}$, $\phi^y_{i}$, and $\phi^z_{i}$ for all $j=1...N_y+1, k=1...N_z+1$. 
Note, that $A_x=B_x \equiv 1$ in the case $u=\phi^x$. 
The end points $i=1$ and $i=N_x+1$ are computed for book-keeping purposes.
Similarly, the operators $B_y$ and $B_z$ do not automatically
include information on the end points at $j=1, N_y+1$ and $k=1, N_z+1$, respectively.
In addition, there are different boundary conditions, namely periodic, Dirichlet and Neumann
boundary conditions, that require an adaption of the matrix elements on the first and last row \cite{NumRec}.

Another complication is that the physical boundary conditions that apply for the vectors $u^{(n)}$ and $u^{(n+1)}$ in equation
(\ref{eqn:solveTridiag}) do not necessarily apply for the intermediate results $u^{(n+1/3)}$ and $u^{(n+2/3)}$. When,
for example, solving the second equation of the system (\ref{eqn:solveTridiag}),
\begin{equation}
A_y u^{(n+2/3)}=u^{(n+1/3)},
\specialeqnum{\ref{eqn:solveTridiag}$^\prime$}
\end{equation}
a correct treatment of the boundary condition for $u^{(n+2/3)}=A_zu^{(n+1)}$ must be implemented into $A_y$.
The matrices $A_x$, $A_y$, and $A_z$ commute as they act in different directions. It is therefore advisable to solve
(\ref{eqn:solveTridiag}) starting in the direction with the simplest boundary condition. For a periodic boundary condition,
for example, the relations $u_{1}=u_{N}$ and $u_{N+1}=u_{2}$ hold at all time levels, including `fractional' ones.

The boundary conditions depend on the geometry of the problem. We choose a system
with a periodic boundary conditions in the $z$-direction. At the interfaces in the $x$- and the $y$-direction, boundary conditions 
for the magnetic field and the order parameter are applied. 
For the order parameter, $\psi_{i,j,k}$, conditions are needed for all values at the faces of the three-dimensional box. 
The grid representation of the periodic boundary condition reads
\begin{equation}
\begin{array}{llll}
\psi_{i,j,1}&=~\psi_{i,j,N_z}~,&~\psi_{i,j,N_z+1}&=~\psi_{i,j,2}\\
\phi^x_{i,j,1}&=~\phi^x_{i,j,N_z}~,&~\phi^x_{i,j,N_z+1}&=~\phi^x_{i,j,2}~.
\end{array}
\label{eqn:Boundary1}
\end{equation}
In the $y$ and $z$ direction we set the supercurrent across the boundary to zero \cite{Grop1996}, i. e.,
\begin{equation}
\begin{array}{ll}
\psi_{1,j,k}&=~\psi_{2,j,k}\exp(-{\rm i}\phi^x_{1,j,k})\\
\psi_{N_x+1,j,k}&=~\psi_{N_x,j,k}\exp(+{\rm i}\phi^x_{N_x,j,k})\\
\psi_{i,1,k}&=~\psi_{i,2,k}\exp(-{\rm i}\phi^y_{i,1,k})\\
\psi_{i,N_y+1,k}&=~\psi_{i,N_y,k}\exp(+{\rm i}\phi^y_{i,N_y,k})~.
\end{array}
\label{eqn:Boundary2}
\end{equation}
Expressions for the end points of the link variables can be found by incorporating information of the magnetic field at the boundaries
of the box.
The three components of the magnetic field are given by the following 
second order finite-difference approximations.
\begin{eqnarray}
\nonumber
B^x_{i,j,k}&=&\frac{1}{h_y h_z} (\phi^y_{i,j,k}-\phi^y_{i,j,k+1}-\phi^z_{i,j,k}+\phi^z_{i,j+1,k})\\
B^y_{i,j,k}&=&\frac{1}{h_z h_x} (\phi^z_{i,j,k}-\phi^z_{i+1,j,k}-\phi^x_{i,j,k}+\phi^x_{i,j,k+1})
\label{eqn:BField}\\
\nonumber
B^z_{i,j,k}&=&\frac{1}{h_x h_y} (\phi^x_{i,j,k}-\phi^x_{i,j+1,k}-\phi^y_{i,j,k}+\phi^y_{i+1,j,k})~.
\end{eqnarray}
From these expressions, appropriate boundary conditions can be obtained. For example, the field $\phi^x_{i,j,k}$ is 
unknown at $j=1$, and we use the last equation to relate the values of $\phi^x_{i,1,k}$ to known values
\begin{equation}
\phi^x_{i,1,k}=-B^z_{i,1,k} h_x h_y+\phi^x_{i,2,k}+\phi^y_{i,1,k}-\phi^y_{i+1,1,k}~.
\label{eqn:Boundary3}
\end{equation}
Equation (\ref{eqn:newFinDiff1}), (\ref{eqn:newFinDiff2}) and (\ref{eqn:solveTridiag})
combined with the boundary conditions (\ref{eqn:Boundary1}), (\ref{eqn:Boundary2}), and (\ref{eqn:Boundary3})
provide all the information needed to solve a three-dimensional problem.

\section{Example 1: Wire with longitudinal field}
As an example, we model an infinite cylindrical wire in three dimensions with an external magnetic field, $H_z$, applied 
along its axis. Such a configuration has been studied in \cite{Blac2000,Gene1995}.
Experiments have shown that the magnetic field can increase the critical current down the wire
\cite{Belo1967}.
\begin{figure}[ht]
\center
\epsfig{file=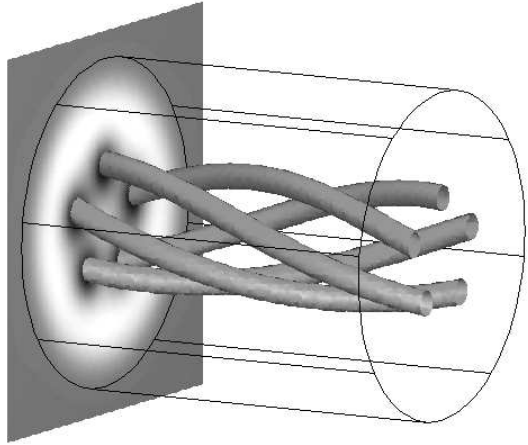,clip=,width=7cm}
\vskip12pt
\label{fig:spiral}
\caption{Time-independent arrangement of spiral vortices in a cylinder of radius 12$\xi$, $\kappa=5$. Five 
vortex lines are entangled with two rings. The applied fields
are $H_z=0.2$ and $H_{\varphi}=1.5/\rho$, respectively. The tubes show the density of the order at the level $|\psi|^2=0.3$. 
The left slice shows the magnetic field component parallel to the wire, $H_z$, dark regions indicate a high field. The black
lines mark the boundaries of the wire.}
\end{figure}
\par In a type-II superconductor, a sufficiently strong magnetic field, $H_z$, will enter the cylinder and become trapped
in vortex tubes aligned parallel to the axis of the cylinder.
A current $I$ along the wire induces an additional circular field $H_{\varphi}(\vek{r})$ such that 
$I=2\pi\rho\kappa^2 H_{\varphi}(\rho)$ at a distance $\rho$ from the axis of the wire. 
The vortex tubes corresponding to the current-
induced $H_{\varphi}$ field are rings coaxial with the wire. Above a critical current, these vortex rings enter 
at the edge of the cylinder
and shrink until they annihilate on the axis of the cylinder. This process repeats and leads to
dissipation. There is no stable mixed state associated with a $H_{\varphi}$ field unless the vortex rings are pinned 
by impurities in the material. Blackburn et.~al.~\cite{Blac2000} have argued, that by entangling the vortex rings with vortex
lines due to a strong longitudinal field, the rings can be prevented from shrinking and thereby increase the critical current 
in the wire.

We model a cylindrical shape of the wire in a rectangular box by adding a potential term $V\psi$ with
$V=5$ to equation (\ref{eqn:TDGL1}) at all grid points outside a cylindrical region with radius $R=12\xi$. The density of the order 
parameter
outside the cylinder decreases rapidly to zero. An array of longitudinal vortex tubes is created by imposing boundary conditions
for an external magnetic parallel to the wire. A current is ramped up by slowly increasing a circular field 
($B_x$ and $B_y$) around the box until vortices enter. The Bean-Livingston
surface barrier was lowered by adding a weak sinusoidal potential at the surface of the cylinder. Fig. 
\ref{fig:spiral} shows a time-independent state that arises after two vortex rings have entered the cylinder and 
entangled with the vortex lines. The critical current is dominated by the surface barrier as expected
for small samples. Consequently, we do not observe any improvement in the critical current due to the presence of a longitudinal field.
However, the effect could become more significant for larger sample sizes or if the surface effects are suppressed
\cite{Wini2001}.

\section{Example 2: Wire with transverse field}
\begin{figure}[ht]
\center
\epsfig{file=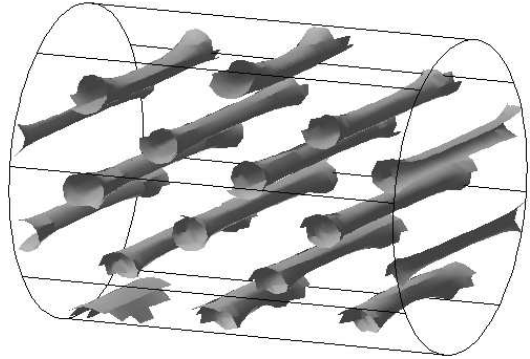,clip=,width=7cm}
\vskip12pt
\caption{Array of vortices in a wire exposed to a perpendicular magnetic field $B_x=0.4$. The vortices tubes 
enter end exit the surface of the superconductor normally. The Ginzburg-Landau parameter is $\kappa=5$, the radius
of the wire $R=8\xi$.}
\label{fig:wireInMagnet}
\end{figure}

In superconducting magnets, the external magnetic field is typically 
aligned perpendicular to the wire ($B_x$ for example). In the mixed state,
an array of vortex lines fills up the superconductor (see Fig. (\ref{fig:wireInMagnet})). Any current carried by the wire superimposes 
a circular field onto this applied field. As a result, a gradient of the magnetic field develops that can be 
associated with a Lorentz force on the vortices. In most applications, different pinning mechanisms balance
this Lorentz force and freeze the flux lattice up to a critical current density. For larger currents,
the Lorentz force exceeds the pinning force and vortices start moving \cite{Wini2001}. The motion of the flux
lattice coincides with the breakdown of superconductivity.
\par In this geometry the magnetic field at the boundary of the computational box strongly depends on the 
currents inside unless the box size is much larger than the radius $R$ of the cylinder. 
In the Meissner state, for example, no flux lines penetrate the wire and supercurrents 
in the surface of the superconductor cancel the external field in the bulk of the wire. 
The field lines of these surface currents also extend outside the sample at length scales 
of order $R$. In the mixed state, the induced field is smaller and can be regarded as a small correction.
To find a self-consistent solution, the fields induced by both the supercurrents and normal currents have to be
added to the applied field and included in the boundary conditions at each time step. We calculate
the induced field $\vek{H}_{\rm ind}(\vek{r})$ using the Biot-Savart law, which in our units has the form,
\begin{equation}
\vek{H}_{\rm ind}(\vek{r})=\frac{1}{4 \pi \kappa^2}\int {\rm d}^3r' \vek{j}(\vek{r}') \times \frac{\vek{r}-\vek{r}'}{|\vek{r}-\vek{r}'|^3}~.
\label{eqn:Biot-Savart}
\end{equation}
This calculation is computationally expensive. The integral is approximated by summing over all grid points
for each boundary point requiring a total of ${\cal O}(N^5)$ calculations, whereas a time step takes 
${\cal O}(N^3)$ calculations for a box of $N^3$ grid points. With periodic boundary conditions, 
the integration must also be extended to regions outside the box. 
The effect of including $\vek{H}_{\rm ind}$ is to bend the vortex lines, especially near the top and the bottom of the sample as apparent in 
Fig \ref{fig:wireInMagnet}.
To model the motion of flux lines, a fully 
self-consistent time-dependent solution can be found by iterating the boundary conditions at each time-step. 
However, in practice the study of the motion of vortices above the critical current does not seem to be feasible. Possible
ways around this problem are to increase the box size so that the induced currents can be neglected, 
to cut-off the integral in (\ref{eqn:Biot-Savart}) at a certain distance from the boundary,
$|\vek{r}-\vek{r}'|<{\cal R}$, or to update $\vek{H}_{\rm ind}(\vek{r})$ only in larger intervals rather than every time step.
The latter approach is especially suitable for finding time-independent solutions.
\section{Concluding remarks}
In summary, we have demonstrated the use of a semi-implicit finite-difference scheme to solve the three-dimensional
time-dependent Ginzburg-Landau equations. The method converges if a Crank-Nicholson scheme is applied to all 
Laplacian-type terms while all other terms are treated explicitly. 
We iterate each time step to achieve second order
accuracy in time. For intermediate values of the Ginzburg-Landau parameter $\kappa$, the method is stable 
and accurate for time-steps two orders of magnitude larger 
than used in the standard explicit schemes. If the magnetic field at the 
surface of the computational box is partly due to currents inside, a self-consistent solution 
can be found by iteration. However, a full Biot-Savart calculation of the induced magnetic field remains
computationally expensive.
%% End of article:

%% optional
% Appendixes

% Appendix without title:
%\appendix{}

% Appendix with title:
%\appendix{Title}

% Appendix with letter:
%\appendix{B}

% Appendix with letter and title:
%\appendix{C}
%\appendixtitle{This is an appendix title}

%% optional
\begin{acknowledgment}
We thank EPSRC for financial support to this project.
\end{acknowledgment}

%% not optional:

%% This command is necessary! ==>>
\end{article} 
\end{document}